\documentclass[conference,9pt]{IEEEtran}
\IEEEoverridecommandlockouts
\usepackage{cite}
\usepackage{amsmath,amssymb,amsfonts}
\usepackage{algorithmic}
\usepackage{graphicx}
\usepackage{textcomp}
\def\BibTeX{{\rm B\kern-.05em{\sc i\kern-.025em b}\kern-.08em
    T\kern-.1667em\lower.7ex\hbox{E}\kern-.125emX}}
    
\usepackage[table]{xcolor}
\usepackage{xcolor}
\usepackage{booktabs}
\usepackage{multirow}
\usepackage{array}
\usepackage{svg}

\usepackage{hyperref}
\usepackage{CJKutf8}
\usepackage[english]{babel}
\usepackage{devanagari}
\usepackage{textcomp}
\usepackage{tipa}
\usepackage{arydshln}

\usepackage[most]{tcolorbox}
\usepackage{float}
\usepackage{xspace}
\tcbset{
  aibox/.style={
    width=230.18663pt,
    top=10pt,
    colback=gray!15,
    colframe=black,
    colbacktitle=black,
    enhanced,
    center,
    attach boxed title to top left={yshift=-0.1in,xshift=0.15in},
    boxed title style={boxrule=0pt,colframe=white,},
  }
}
\newtcolorbox{AIbox}[2][]{aibox,title=#2,#1}
    
\begin{document}

\title{MacST: Multi-Accent Speech Synthesis via Text Transliteration for Accent Conversion%\\
%{\footnotesize \textsuperscript{*}Note: Sub-titles are not captured for https://ieeexplore.ieee.org  and should not be used}
\thanks{Work was done when Sho Inoue was doing an internship at NetEase}
\thanks{Sho Inoue: shoinoue@link.cuhk.edu.cn}
\thanks{$\dag$ Correspondence to Shuai Wang: wangshuai@cuhk.edu.cn}
% \thanks{The research is supported by National Natural Science Foundation of China (Grant No. 62271432); Internal Project of Shenzhen Research Institute of Big Data (Grant No. T00120220002); Shenzhen Science and Technology Program ZDSYS20230626091302006; Shenzhen Science and Technology Research Fund (Fundamental Research Key Project Grant No. JCYJ20220818103001002); and CCF-NetEase ThunderFire Innovation Research Funding (No. CCF-Netease 202302).}
\thanks{The research is supported by National Natural Science Foundation of China (Grant No. 62271432); Shenzhen Science and Technology Program ZDSYS20230626091302006; and CCF-NetEase ThunderFire Innovation Research Funding (No. CCF-Netease 202302).}
}

\author{
\IEEEauthorblockN{Sho Inoue\textsuperscript{1,2,3}, Shuai Wang\textsuperscript{1,2$\dag$}, Wanxing Wang\textsuperscript{3}, Pengcheng Zhu\textsuperscript{3},
Mengxiao Bi\textsuperscript{3},
Haizhou Li\textsuperscript{1,2,4}}
%\\
\IEEEauthorblockA{
\textsuperscript{1}\textit{School of Data Science} 
\textsuperscript{2}\textit{Shenzhen Research Institute of Big Data} \\
\textit{The Chinese University of Hong Kong, Shenzhen (CUHK-Shenzhen), Shenzhen, China} \\
\textsuperscript{3}\textit{Fuxi AI Lab, NetEase Inc., Hangzhou, China} \\
\textsuperscript{4}\textit{Department of Electrical and Computer Engineering, National University of Singapore, Singapore} \\
%\textsuperscript{4}\textit{Machine Learning Lab (MLL), University of Bremen, Bremen, Germany} \\
%shoinoue@link.cuhk.edu.cn
%\{wangshuai$|$haizhouli\}@cuhk.edu.cn; 
%\{wangwanxing$|$zhupengcheng$|$bimengxiao\}@corp.netease.com
}
}

%\author{\IEEEauthorblockN{Sho Inoue}
%\IEEEauthorblockA{
%\textit{School of Data Science, The Chinese University of Hong Kong, Shenzhen (CUHK-Shenzhen), China}\\
%\textit{Shenzhen Research Institute of Big Data, Shenzhen, China} \\
%\textit{NetEase, Hangzhou, China}\\
%email address or ORCID
%}
%\and
%\IEEEauthorblockN{Shuai Wang\textsuperscript{$\dag$}}
%\IEEEauthorblockA{
%\textit{Shenzhen Research Institute of Big Data, Shenzhen, China} \\
%email address or ORCID
%}
%\and
%\IEEEauthorblockN{Wanxing Wang}
%\IEEEauthorblockA{
%\textit{NetEase, Hangzhou, China}\\
%email address or ORCID
%}
%\and
%\IEEEauthorblockN{Zhu Pengcheng}
%\IEEEauthorblockA{
%\textit{NetEase, Hangzhou, China}\\
%email address or ORCID
%}
%\and
%\IEEEauthorblockN{Haizhou Li}
%\IEEEauthorblockA{
%\textit{School of Data Science, The Chinese University of Hong Kong, Shenzhen (CUHK-Shenzhen), China}\\
%\textit{Shenzhen Research Institute of Big Data, Shenzhen, China} \\
%email address or ORCID
%}
%}

\maketitle

\begin{abstract}

In accented voice conversion or accent conversion, we seek to convert the accent in speech from one another while preserving speaker identity and semantic content. In this study, we formulate a novel method for creating multi-accented speech samples, thus pairs of accented speech samples by the same speaker, through text transliteration for training accent conversion systems. We begin by generating transliterated text with Large Language Models (LLMs), which is then fed into multilingual TTS models to synthesize accented English speech. As a reference system, we built a sequence-to-sequence model on the synthetic parallel corpus for accent conversion. We validated the proposed method for both native and non-native English speakers. Subjective and objective evaluations further validate our dataset's effectiveness in accent conversion studies. 
%\textcolor{red}{(to Sho: I don't understand why the method is `scalable'? i revise the title. please double check and )}

\end{abstract}

\section{Introduction}\label{sec:intro}

Despite much progress in expressive speech generation, many challenges remain in accent speech generation. Supervised learning effectively aligns phonetic and prosodic features across accents that relies on parallel corpus. One of the major challenges is the scarcity of accent-varied parallel speech corpus since multi-accent speakers are hard to come by~\cite{AccentVITS}. 
Therefore, it is common to create parallel corpus by synthesizing target speech using Voice Conversion (VC) ~\cite{accentevaluatingmethods,VoiceShop,accentvcphone,accentttsfrontend} or text-to-speech (TTS)~\cite{accentvoicecloning}. However, these techniques also depend on accented speech corpus and often face speaker entanglement issues.

Recent advancements in TTS technologies and Large Language Models (LLMs) open up new possibilities. Multi-lingual TTS systems have achieved significant progress, now producing speech that closely mirrors human-like naturalness across various languages~\cite{XTTS,YourTTS,MegaTTS2,VALLEX}. In parallel, LLMs have revolutionized text generation tasks. Initially, generating high-quality text was a time-consuming process and required specialized knowledge~\cite{Transliteration1,Transliteration2}. However, with the advent of LLMs~\cite{LLM1,GPT4}, these tasks have become more efficient and accessible. We apply these developments to generate a parallel dataset for accent conversion.

In this study, we introduce a novel method for generating a multi-accent speech samples via text transliteration. In practice, we first convert text from one language to another while maintaining the phonetic equivalence. Table~\ref{tab:transliteration} illustrates transliterated examples of the word ``accent'' across three languages. The transliteration is done by Large Language Models (LLMs). The transliterated text is  subsequently taken by a multilingual TTS model to synthesize the accented English speech. As phonetic variation represents the main signature of an accent~\cite{LexicalProsody2,Lexical1}, this process allows us to effectively construct a parallel accent dataset that varies solely in accent.

\vskip-1.2em
\begin{table}[ht]
\centering
\caption{English word ``accent'' and its transliterations}
\scalebox{0.9}{
\begin{tabular}{ccc}
\toprule
Language & Transliteration (``Accent'') & Pronunciation \\
\midrule
Hindi & {\dn akseMT} & aksemt \\
Japanese & \begin{CJK}{UTF8}{min}アクセント\end{CJK} & akusento \\
Korean & \begin{CJK}{UTF8}{mj}액센트\end{CJK} & aegsenteu \\
\bottomrule
\end{tabular}
}
%\vskip-0.7em
\label{tab:transliteration}
\end{table}

This study is motivated to create accented English speech without the need to involve human speakers, therefore avoiding the issue of English proficiency of speakers~\cite{L2ARCTIC}. Unlike other traditional paired data generation methods, such as speaker voice conversion (VC), the multi-accent speech synthesis via text transliteration method, i.e. MacST, offers unique benefits:
%(1) \textit{Phonetic variation via transliteration}: MacST directly varies the phonemes across accents without depending on spoken samples, which avoids entanglement between speaker and accent. 
%(2) \textit{Generalization of linguistic content}: Unlike VC-augmented methods, which are restricted to linguistic content from existing speech samples and thus unable to handle low-resource English accents, MacST is applicable to any English sentence.

\begin{itemize}

\item \textit{Phonetic variation via transliteration}: MacST directly varies the phonemes across accents without depending on spoken samples, which avoids entanglement between speaker and accent. %Our approach overcomes the challenges of ensuring that VC models exclusively modify speaker characteristics without affecting other speech features. 

\item \textit{Generalization of linguistic content}: Unlike VC-augmented methods, which are restricted to linguistic content from existing speech samples and thus unable to handle low-resource English accents, MacST is applicable to any English sentence.

%\item \textit{Customizable Accent Variation}: MacST allows for specific adjustments in accent features, such as enabling or disabling particular phonemes. \textcolor{red}{(to Sho: do we test this unique benefit?). I dont explicitly test it in the experiment. This can be done by changing the LLM prompt. But, this can be confusing so I remove this part.}

\end{itemize}
In short, we are seeking a speech generation method that is simple, scalable, and applicable to general accent conversion studies. The  contributions of this work can be summarized as follows:
\begin{itemize}
  \setlength{\leftmargini}{10pt}
  \item We introduce the first approach utilizing transliteration to construct a parallel accent dataset. we can enhance accent intensity by modeling the absence of specific English phonemes in the first language.
  \item Analysis of our dataset confirms the effectiveness of our method in generating accented speech from both native and non-native English speakers, intensifying the latter's accents.
  \item Experimental results demonstrate that our synthetic parallel dataset significantly enhances the performance of accent voice conversion systems.
\end{itemize}

The rest of this paper is organized as follows: In Section II, we introduce related works. Section III describes our proposed methodology. In Section IV, we introduce our experiment setup. In Section V, we validate our work with experimental results and analysis. Section VI concludes our study. 
%Audio samples\footnote{\textbf{Speech Demos}: \url{https://shinshoji01.github.io/MAcST-Demo/}} and the project page\footnote{\textbf{Project Page}: \url{}} are available online.
We placed speech demos, transliterated texts, and training datasets on the project page\footnote{\textbf{Project Page}: \url{https://github.com/shinshoji01/MacST-project-page}}.

\begin{figure*}[ht]
  \centering
  \centerline{\includegraphics[width=15cm]{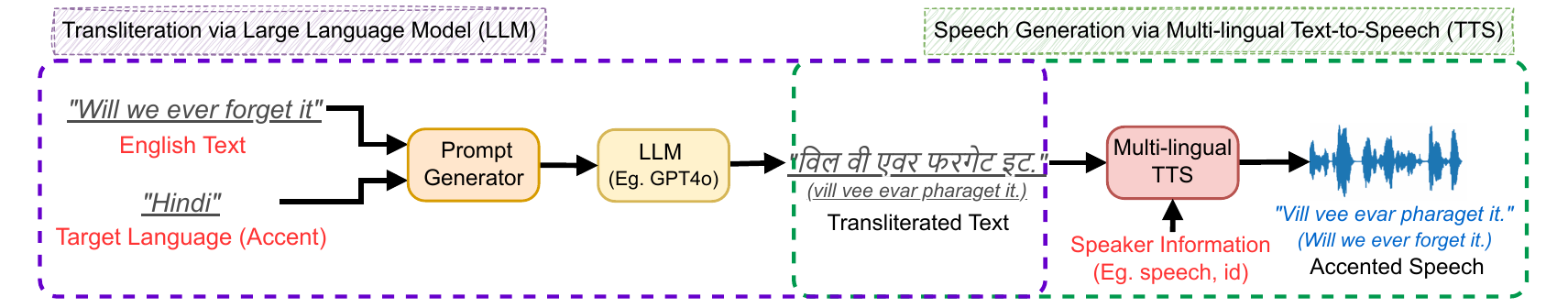}}
  %\vspace{-0.3cm}
 \caption{Overall diagram of the MacST pipeline: The system first generates transliterated text from the input, which is then fed into multi-lingual TTS models to synthesize accented speech. 
 Red texts denote our proposed system's input data (English Text, Target Language, and Speaker Information).
 }
 
%\vspace{-0.4cm}
 \label{fig:diagram}
\end{figure*}

\section{Related Works}\label{sec:related_works}

\subsection{Accent Speech Dataset}

Numerous datasets featuring accented English speech have been made available for various speech processing applications. The Edinburgh Dataset~\cite{EdinburghDataset} offers 40 hours of phone conversation recordings from native English speakers with diverse accents, designed primarily for automatic speech recognition tasks. The VCTK dataset~\cite{VCTK} is tailored for text-to-speech synthesis (TTS) and includes 109 native English speakers from different regions, such as the US and the UK. GLOBE~\cite{GLOBE} is a large-scale multi-speaker TTS dataset that contains over 20,000 speakers and 150 accents, including non-native speakers like Europeans and Asians. Finally, the combination of L2-ARCTIC~\cite{L2ARCTIC} and CMU-ARCTIC~\cite{CMUARCTIC} contain both native and non-native speakers, sharing identical transcriptions to construct parallel datasets. However, \emph{each speaker is restricted to a single accent.}

\subsection{Accent Conversion}
The studies on accent conversion can be summarized into two categories according to the use of speech corpus. % address this issue, the following two primary approaches are employed.

\noindent
%\textbf{(1) Dataset with Synthetic Samples}:
\subsubsection{Synthetic Parallel Corpus}
The first direction is to synthesize audio to create parallel data for accent conversion. They generate native-like speech from non-native English speakers, or accented speech from native speakers via voice conversion (VC)~\cite{accentevaluatingmethods,VoiceShop,accentvcphone,accentttsfrontend}. For example, a study~\cite{accentevaluatingmethods} explores three ground-truth-free methods for accent conversion including synthesizing data from VC. 
%Additionally, \cite{accentvcphone} and \cite{accentttsfrontend} address accent conversion (AC) in low-resource accents by pretraining the linguistic encoder and TTS front-end, respectively.
%Other approaches include training a TTS model only on a specific accent to produce accented speech across various speakers~\cite{accentvoicecloning}. 
%Another study~\cite{accentnoreference} constructs a speech synthesizer with a decoder trained solely on native English, combined with non-native speaker embeddings to generate target speech. 
%\textcolor{blue}{
Other approaches~\cite{accentvoicecloning,accentnoreference} include training speech generation models only on a specific accent to produce accented speech across various speakers. 
However, these methods often encounter speaker entanglement issues and restrict their application to existing accented corpora.
%}

\noindent
%\textbf{(2) Non-Parallel Accent Conversion}:
\subsubsection{Non-Parallel Corpus}
Accent conversion can be achieved using non-parallel datasets. Techniques involve training a decoder solely on the target speaking style, thereby eliminating the need for native accent speech during conversion~\cite{accentnonative1,accentchenxi,accentnonative2}. Accentron employs speaker/accent encoders trained with speaker/accent classification tasks~\cite{accentron}. Also, through adversarial learning, some studies facilitate many-to-many accent conversion in Chinese~\cite{accentmanytomany} and English \cite{accentmeta}. It is generally believed that supervised learning on parallel corpus is also more effective than unsupervised learning on non-parallel corpus.

{
In this paper, we study a novel way to automatically construct accent-parallel speech corpus. This corpus will facilitate the alignment of phonetic and prosodic features across accents, thereby simplifying the training process.
}

%\textcolor{red}{(to Sho: we need to give comments here. If this paper is about generating parallel corpus, we should argue that parallel corpus is better than non-parallel corpus in some way here. )}, OK professor

\section{MacST Method}

\subsection{Overall Pipeline}
We designed a text-to-speech synthesis pipeline that takes text, accent ID, and speaker information as input and generate accented speech as output via accent-transliterated text, as outlined in Fig.~\ref{fig:diagram}. The process comprises two main steps: transliteration through Large Language Models (LLMs) and speech synthesis using a multi-lingual Text-to-Speech (TTS) model. This method notably applies across different speakers, accents, and transcriptions, allowing for extensive speech sample generation.

\begin{figure}[!htb]
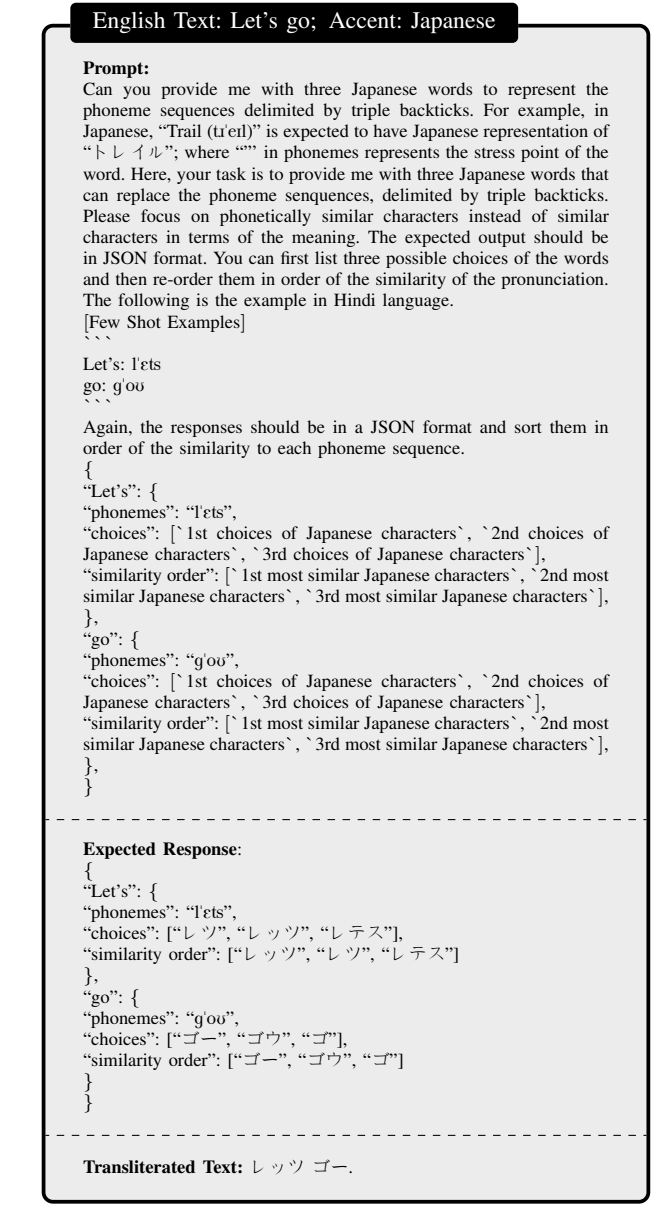

\begin{AIbox}{English Text: Let's go;\hskip 0.5em Accent: Japanese}
%\begin{AIbox}{}
{\scriptsize
%{\bf English Text:} Let's go. \hskip 3em {\bf Accent:} Japanese
%\tcbline
{\bf Prompt:} \\
Can you provide me with three Japanese words to represent the phoneme sequences delimited by triple backticks. For example, in Japanese, ``Trail (\textipa{t\textturnr"eIl})'' is expected to have Japanese representation of ``\begin{CJK}{UTF8}{min}トレイル''\end{CJK}; where ``''' in phonemes represents the stress point of the word. Here, your task is to provide me with three Japanese words that can replace the phoneme sequences, delimited by triple backticks. Please focus on phonetically similar characters instead of similar characters in terms of the meaning. The expected output should be in JSON format. You can first list three possible choices of the words and then re-order them in order of the similarity of the pronunciation. \\
The following is the example in Hindi language.\\
$[$Few Shot Examples$]$\\
\textasciigrave\textasciigrave\textasciigrave\\
Let's: \textipa{l"Ets} \\\
go: \textipa{g"oU}\\
\textasciigrave\textasciigrave\textasciigrave\\
Again, the responses should be in a JSON format and sort them in order of the similarity to each phoneme sequence.\\
\{\\
  ``Let's'': \{\\
``phonemes'': ``\textipa{l"Ets}'',\\
``choices'': $[$\textasciigrave 1st choices of Japanese characters\textasciigrave , \textasciigrave 2nd choices of Japanese characters\textasciigrave , \textasciigrave 3rd choices of Japanese characters\textasciigrave $]$,\\
``similarity order'': $[$\textasciigrave 1st most similar Japanese characters\textasciigrave , \textasciigrave 2nd most similar Japanese characters\textasciigrave , \textasciigrave 3rd most similar Japanese characters\textasciigrave $]$,\\
\},\\
  ``go'': \{\\
``phonemes'': ``\textipa{g"oU}'',\\
``choices'': $[$\textasciigrave 1st choices of Japanese characters\textasciigrave , \textasciigrave 2nd choices of Japanese characters\textasciigrave , \textasciigrave 3rd choices of Japanese characters\textasciigrave $]$,\\
``similarity order'': $[$\textasciigrave 1st most similar Japanese characters\textasciigrave , \textasciigrave 2nd most similar Japanese characters\textasciigrave , \textasciigrave 3rd most similar Japanese characters\textasciigrave $]$,\\
\},\\
\}
\tcbline
{\bf Expected Response}:\\
\{\\
  ``Let's'': \{\\
    ``phonemes'': ``\textipa{l"Ets}'',\\
    ``choices'': [``\begin{CJK}{UTF8}{min}レツ\end{CJK}'', ``\begin{CJK}{UTF8}{min}レッツ\end{CJK}'', ``\begin{CJK}{UTF8}{min}レテス\end{CJK}''],\\
    ``similarity order'': [``\begin{CJK}{UTF8}{min}レッツ\end{CJK}'', ``\begin{CJK}{UTF8}{min}レツ\end{CJK}'', ``\begin{CJK}{UTF8}{min}レテス\end{CJK}'']\\
  \},\\
  ``go'': \{\\
    ``phonemes'': ``\textipa{g"oU}'',\\
    ``choices'': [``\begin{CJK}{UTF8}{min}ゴー\end{CJK}'', ``\begin{CJK}{UTF8}{min}ゴウ\end{CJK}'', ``\begin{CJK}{UTF8}{min}ゴ\end{CJK}''],\\
    ``similarity order'': [``\begin{CJK}{UTF8}{min}ゴー\end{CJK}'', ``\begin{CJK}{UTF8}{min}ゴウ\end{CJK}'', ``\begin{CJK}{UTF8}{min}ゴ\end{CJK}'']\\
  \}\\
\}
\tcbline
{\bf Transliterated Text:} \begin{CJK}{UTF8}{min}レッツ ゴー\end{CJK}.
}
\end{AIbox}
\caption{
%Examples of the transliteration process of the text ``Let's go'': The prompt of LLM and the expected response. Example responses are placed in $[$Few Shot Examples$]$.
Examples of the transliteration process: The prompt of LLM and the expected response. Example responses are placed in $[$Few Shot Examples$]$.
}
\label{fig:llm_prompt}
\end{figure}

\subsection{Transliteration by Large Language Model}
We utilized Large Language Models (LLMs) to obtain the transliterated text from the provided English text. We show the sample prompt and the expected response of the English text ``Let's go'' in Fig.~\ref{fig:llm_prompt}. We constructed a prompt for the LLM to transliterate the English sentence at the word level, associating each word with a phoneme sequence. This method is supported by research suggesting that the inclusion of both graphemes and phonemes enhances transliteration accuracy~\cite{transliterationsurvey}. We designed the prompt to provide three transliteration candidates per word, then sorted by similarity. We include a few transliterated samples to prevent LLM from \textit{translating} samples. We executed this prompt six times, 
%three with GPT-3.5 Turbo\footnote{gpt-3.5-turbo-1106: \url{https://platform.openai.com/docs/models}} and three with GPT-4o\footnote{gpt-4o-2024-05-13: \url{https://platform.openai.com/docs/models}}. 
three with GPT-3.5 Turbo~\cite{GPT3.5Turbo} and three with GPT-4o~\cite{GPT4o}.
We calculated the frequency of each transliterated word, assigning higher scores to those with similar representations. 
Concurrently, we obtained transliterations for articles like ``the'' and ``a/an'' since their pronunciations vary depending on the subsequent word. We then concatenated the top-scoring transliterations, adding commas and periods to form complete sentences representing the original English text.

%- post processing 
%    - pronunciation change in different sentences
%        - ``a'': "ah" instead of "ei"
%        - ``the'': "thi" or "tha"
%    - It keeps "," and "." to have similar pause as English.
%    - put higher scores on more similar phrases.

\subsection{Speech Generation via Multi-lingual TTS}

We integrate a multi-lingual Text-to-Speech (TTS) system capable of handling multiple target languages. This system, provided by 11Elevenlabs\footnote{11Elevenlabs: \url{https://elevenlabs.io/}}, employs the Eleven Multilingual v2 model, which supports 29 languages. It generates speech from the transliterated text and speaker information, producing accented English speech specific to the chosen speaker. The model is conditioned on the speaker using audio recordings from a speaker and conditioned on language using transcription rather than a language ID. Therefore, our approach is effective for languages with characters distinct from English, such as Hindi, Korean, Japanese, and Mandarin.
%\textcolor{red}{(to Sho: This paper claims to propose a method to generate accented corpus. In the experiments, it seems that we use accented parallel data to train a sequence -to -sequence model. However, in the paper itself, we never mentioned how many utterances are generated? are we generating parallel corpus?)}, OK professor. I will put this content in Section V-A and V-B.

\section{Experiment Setup}

%\textcolor{red}{(to Sho: we add a line here to explain the purpose of the experiement - We develop a sequence-to-sequence voice conversion model based on the accented parallel corpus that we generate with MAcST.}, OK professor

{
We developed a sequence-to-sequence voice conversion model on an accented parallel corpus generated by MacST. 
}

\subsection{Voice Conversion Model Configuration}
Following~\cite{accentevaluatingmethods}, we employ Voice Transformer Network (VTN)~\cite{VTN} as a sequence-to-sequence model for accent voice conversion.  This model adopts a standard encoder-decoder architecture comprising 12 Transformer encoder blocks and 6 decoder blocks~\cite{Transformer}, with model dimensions set at 768, feed-forward network (FFN) inner dimensions at 3,072, and 12 attention heads. 
We used mel-spectrograms for both input and output acoustic features. 
%For waveform synthesis, we trained the Vocos\footnote{Vocos: \url{https://github.com/gemelo-ai/vocos}}~\cite{Vocos} vocoder on our internal speech dataset, which was sampled at 16kHz and featured a 100-dimensional mel-spectrogram, aligning with the CMU-ARCTIC dataset. 
For waveform synthesis, we trained the HiFiGAN\footnote{HiFiGAN: \url{https://github.com/jik876/hifi-gan}}~\cite{HiFiGAN} vocoder on LibriTTS-R~\cite{LibriTTSR} and ARCTIC datasets~\cite{L2ARCTIC,CMUARCTIC}, which was sampled at 16kHz and featured a 80-dimensional mel-spectrogram, aligning with CMU-ARCTIC. 

To enhance training stability, we implemented a two-stage pretraining strategy from VTN~\cite{VTN}, focusing sequentially on the decoder and encoder. Initially, the model learns to convert linguistic representations into mel-spectrograms. Unlike the text tokens used in~\cite{VTN}, we used Hubert discrete tokens~\cite{Hubert}, extracting continuous features from Hubert Base\footnote{Hubert Base: \url{https://huggingface.co/facebook/hubert-base-ls960}}, which we then discretize using k-means clustering into 500 clusters, eliminating repetitive tokens. 
%In the second stage, we replace the input with a mel-spectrogram, maintaining the decoder's parameters from the initial stage and freezing them to train only the encoder. 
%\textcolor{blue}{
In the second stage, we replaced the input with a mel-spectrogram, while maintaining and freezing the decoder's parameters from the initial stage and training only the encoder.
%}
Then, for accent conversion, we initialized the system with parameters from the second stage and fine-tuned using a parallel dataset to adapt to different accents.

We trained the model across three stages: initially, the pretraining phase involved 200,000 steps; followed by a second pretraining stage of 50,000 steps; and finally, the accent conversion stage, comprising 100,000 steps. Batch sizes were set at 8 for the first stage and 64 for the latter two. Apart from these modifications, we adhered to the same training configuration as VTN\footnote{VTN: \url{https://github.com/unilight/seq2seq-vc}}.

\subsection{Dataset}
%\textcolor{red}{(to Sho: This subsection needs to re-write. it mixed up several databases, and did not explain a purpose?)}
%\textcolor{red}{(to Sho: do we generate accent pairs?)}
%\textcolor{red}{(to Sho: if L2-ARCTIC already has parallel accent pairs, our MacST also generates pairs, what is the use of L2-ARCTIC?}
%\textcolor{red}{(to Sho: I am guessing that in the Dataset section, you first talk about using MacST to generate a corpus, however, there is no information about the generated corpus - how many utterances, how many speakers ... if you specify them in experiments, at least you need to say that we would like to create a database that is for accent conversion system training.}
%\textcolor{red}{(to Sho: what is the use of the additional datasets? additional to what?)}

%\textcolor{red}{(I suppose that we only use the text script of ARCTIC? not the speech samples.)
%}
%\textcolor{orange}{
%From SHO :::\\
%For dataset analysis, I used the audio samples of L2-ARCTIC and CMU-ARCTIC: 4 speakers from L2-ARCTIC and 1 speaker from CMU-ARCTIC.\\
%For accent conversion, I used the speech samples of 1 speaker from CMU-ARCTIC. The speech samples are used as source audio of the accent conversion and its target audio is synthesized by MacST.\\
%Also, all speakers' speech samples are used as speaker prompts for multi-lingual TTS.
%I'll mention this one in Section V
%}

Four speech datasets were involved in this paper. L2-ARCTIC~\cite{L2ARCTIC} and CMU-ARCTIC~\cite{CMUARCTIC} were employed to generate samples from MacST for dataset analysis. For accent voice conversion experiments, we used CMU-ARCTIC, LibriTTS-R~\cite{LibriTTSR}, and VCTK~\cite{VCTK}.

L2-ARCTIC~\cite{L2ARCTIC} includes English speech recordings from 24 non-native speakers with six backgrounds such as Hindi and Korean. All speakers share the same script, each offering one hour of speech in a single accent. We also employed CMU-ARCTIC~\cite{CMUARCTIC}, a native version of L2-ARCTIC. For accent conversion, we selected an American speaker from CMU-ARCTIC and divided 1,132 transcriptions into 932 for training, 100 for validation, and 100 for testing. Utilizing both ARCTIC datasets, we created speech samples via MacST for dataset analysis and to build a parallel dataset, as outlined in Section~\ref{sec-experiments}.

In our accent voice conversion experiments, we utilized several datasets including CMU-ARCTIC, LibriTTS-R~\cite{LibriTTSR} and VCTK~\cite{VCTK}. For pre-training, we used LibriTTS-R~\cite{LibriTTSR}, a multi-speaker dataset with approximately 580 hours from 2,306 speakers, specifically employing ``train-clean-100'' and ``train-clean-360'' subsets for training.  For training conversion models, we mainly used the parallel dataset constructed by MacST from CMU-ARCTIC. Additionally, we used transcriptions from VCTK~\cite{VCTK} to test data augmentation, extracting transcriptions with fewer than 15 words and selecting 4500 of them (around 3 hours) to synthesize pairs of American and Hindi accented speech via MacST. This confirmed the usability of our synthetic data in data augmentation practices.

%We generated speech samples by MacST based on ARCTIC datasets~\cite{L2ARCTIC,CMUARCTIC}. The L2-ARCTIC dataset~\cite{L2ARCTIC} comprises parallel English speech recordings from 24 non-native speakers of backgrounds such as Hindi, Korean, Mandarin, Spanish, Arabic, and Vietnamese, with each providing approximately one hour of identical transcriptions. \textcolor{red}{(to Sho: if L2-ARCTIC already has parallel accent pairs, our MacST also generates pairs, what is the use of L2-ARCTIC?} Additionally, we utilized the CMU-ARCTIC dataset~\cite{CMUARCTIC}, the native version of L2-ARCTIC. We selected an American speaker for accent conversion and four L2-ARCTIC speakers for dataset analysis. We distributed 1132 transcriptions into 932, 100, and 100 samples for training, validation, and testing, respectively. \textcolor{red}{(to Sho: I don't understand the purpose of this paragraph.)}

\begin{table*}[!ht]
\caption{
%\textcolor{blue}{
Results for Accent Conversion (AC):
AC transforms American-accented speech into Hindi-accented speech using speaker ``SLT''.
%}
}
\label{table:result_conversion}
\centering
\scalebox{0.9}{
\begin{tabular}{lcccccc}
\toprule
 & \multicolumn{2}{c}{Speech Quality} & \multicolumn{3}{c}{Accentedness} & \multicolumn{1}{c}{Speaker Similarity}\\
\cmidrule(lr){2-3}\cmidrule(lr){4-6}\cmidrule(lr){7-7}
  & MUSHRA ($\uparrow$) & WER ($\downarrow$) & MUSHRA ($\uparrow$) & Classification Prob. ($\uparrow$) & AECS Diff. ($\uparrow$) & SECS ($\uparrow$)\\
\midrule
Ground-Truth (American) &76.48{\tiny $\pm$ {3.82} } & 1.97 & 9.56{\tiny $\pm$ {1.32} } & 0.000 & - & -\\
MacST (American) &70.95{\tiny $\pm$ {4.07} } & 1.75 & 10.78{\tiny $\pm$ {1.41} } & 0.000 & - & 0.866\\
MacST (Hindi) &69.51{\tiny $\pm$ {3.99} } & 8.52 & 51.61{\tiny $\pm$ {3.02} } & 0.819 & - & 0.822\\
\midrule
AC w/o Data Augmentation &51.48{\tiny $\pm$ {3.73} } & 13.99 & 34.85{\tiny $\pm$ {2.29} } & 0.801 & 0.411 & \textbf{0.834}\\
AC w/ Data Augmentation (ours) &\textbf{67.18}{\tiny $\pm$ {3.43} } & \textbf{8.74} & \textbf{47.26}{\tiny $\pm$ {2.65} } & \textbf{0.897} & \textbf{0.465} & 0.833\\
\bottomrule
\end{tabular}
}
\end{table*}

\subsection{Evaluation Metrics}

We evaluated our results against two metrics: speech quality and accentedness. We conducted listening tests with 20 evaluators, each 10 participants familiar with either Korean or Hindi accents.

\noindent
\textbf{(1) Speech Quality}:
%\subsubsection{Speech Quality}
We evaluated speech quality through both subjective and objective methods. Subjectively, we conducted a MUSHRA test in which evaluators rated each audio sample on a scale from 0 to 100, focusing on whether the speech sounded as if it were spoken by humans, explicitly disregarding accents or background noise. Objectively, we used Word Error Rate (WER), employing Whisper\footnote{Whisper Large: \url{https://github.com/openai/whisper}}~\cite{whisper} to predict the transcription of the synthesized audio. %It is important to note that for accented speech, lower metric values do not necessarily indicate superior performance.

\noindent
\textbf{(2) Accentedness}:
%\subsubsection{Accentedness}
Accentedness assesses the prominence of the accent in speech. Subjectively, we conducted a MUSHRA test, where evaluators rated each audio sample based on the strength of the accent. Objectively, we utilized a pretrained accent detector\footnote{\url{https://huggingface.co/Jzuluaga/accent-id-commonaccent\_xlsr-en-english}}~\cite{CommonAccent} to determine the classification probability of Hindi accents. We evaluated the effectiveness of accent conversion using synthetic speech from MacST. Using the pretrained accent classifier, we extracted accent embeddings from three samples: converted speech from the accent conversion process, accented samples, and American samples from MacST. We computed the cosine similarity for accent embeddings (AECS) of the accented and the American speech samples toward the converted speech. We then analyzed the difference of these similarities, such as $(\text{AECS}_{\text{accented}}-\text{AECS}_{\text{native}})$. A higher AECS difference indicates better accent alignment between the converted speech and the accented speech.

\noindent
\textbf{(3) Speaker Similarity}:
%\textcolor{blue}{
We evaluated the speaker perseverance of our conversion models using Speaker Encoding Cosine Similarity (SECS), employing Resemblyzer\footnote{Resemblyzer: \url{https://github.com/resemble-ai/Resemblyzer}} for speaker embedding extraction. We computed SECS using ground-truth source audio from the American speaker (SLT).
%}

%\noindent
%\textbf{(3) Speaker and Accent Similarity}:
%\subsubsection{Speaker and Accent Similarity}
%We evaluated the effectiveness of accent addition using synthetic speech from MAcST. We extracted accent and speaker embeddings from converted, accented, and native samples using pre-trained accent classifiers and speaker verification models. Specifically, we used the same classifier for accent evaluation and three models for speaker verification: WeSpeaker\footnote{WeSpeaker: \url{https://github.com/wenet-e2e/wespeaker}}~\cite{wespeaker}, WavLM Base with X-vector\footnote{WavLM Base with X-vector: \url{https://huggingface.co/microsoft/wavlm-base-plus-sv}}~\cite{wavlm}, and Resemblyzer\footnote{Resemblyzer: \url{https://github.com/resemble-ai/Resemblyzer}}. We computed the cosine similarity for accent and speaker embeddings (AECS and SECS, respectively) of the accented and native speech samples toward the converted speech. We then analyzed the ratios of these similarities, such as $\frac{\text{SECS}_{\text{accented}}}{\text{SECS}_{\text{native}}}$. A higher AECS ratio indicates better accent alignment in the converted speech, while an SECS ratio close to 1 suggests consistent speaker identity.

%\subsubsection{Prosodic Similarity to Synthetic Ground Truth}
%We evaluate the accent/speech quality of accent-converted audio by comparing the converted audio with the synthetic target speech. We calculated four objective scores: (1) Melcepstral Distortion (MCD)~\cite{mcd} for spectral similarity, (2,3) Pitch and Energy Distortion for prosody alignment, and (4) Frame Disturbance (FD)~\cite{fd} for duration deviation. 

\section{Experiments and Results}\label{sec-experiments}

\subsection{MacST: Dataset Analysis}

We assessed the quality of generated speech from MacST against existing datasets. We focused on four speakers from L2-ARCTIC: \textit{ASI} (Hindi male), \textit{TNI} (Hindi female), \textit{HKK} (Korean male), and \textit{YDCK} (Korean female), along with \textit{SLT} (American female) from CMU-ARCTIC. We generated speech samples in seven styles using MacST, employing 100 transcriptions from the ARCTIC datasets' test split. 
We used MacST on non-native speakers (\textit{ASI}, \textit{TNI}, \textit{HKK}, \textit{YDCK}) to test accent enhancement abilities by modifying their linguistic content through transliteration.  We also synthesized accented speeches in American, Hindi, and Korean using speech prompts from the American speaker (\textit{SLT}) to evaluate our method's accented speech generation capabilities. Such speech prompts provide the speaker condition for the multi-lingual TTS system.
% \textcolor{blue}{
% In this paper, the multi-lingual TTS system, from 11Elevenlabs, utilized speech samples as a speaker condition.
% }
%Initially, we synthesized three accented speeches using \textit{SLT} to test the capabilities of accented speech generation by conditioning multi-lingual TTS on the \textit{SLT} speaker's speech prompts. Subsequently, we applied our method to the accented speakers (\textit{ASI}, \textit{TNI}, \textit{HKK}, \textit{YDCK}) to test accent enhancement abilities by modifying their linguistic content through transliteration.

Table~\ref{table:result_analysis} displays MUSHRA tests in terms of speech naturalness and accentedness from two listening groups familiar with Hindi and Korean accents in the upper and the lower sections, respectively. In MacST, the languages in brackets indicate the transliteration languages. Notably, accented speakers with transliterated texts, such as ``MacST (ASI/Hindi)'' and ``MacST (HKK/Korean)'', outperform other cases, suggesting MacST's efficacy in accentuating non-native speakers' accents. When comparing to native English speaker (SLT) results, transliterated texts show heightened accentedness; In the Hindi group, it is hightened from 9.56 to 51.61 and in the Korean group, it is from 6.90 to 77.63. It underscores MacST's ability to accentuate even the native English speaker.

In the Hindi accent group, speech naturalness is consistent across speakers. However, in the Korean group, we observed a slight degradation in speech naturalness, likely due to overly intense accentuation, as indicated by the accentedness score, which is twice as high. To mitigate this, we could reduce the accent intensity, potentially by incorporating some English phonemes in multi-lingual speech generation with a universal multi-lingual tokenizer, for instance. 

\begin{table}[!ht]
\caption{MUSHRA results in terms of naturalness and accentedness across accented corpora.}
\label{table:result_analysis}
\centering
\scalebox{0.90}{
\begin{tabular}{lcc}
\toprule
  & Naturalness ($\uparrow$) & Accentedness ($\uparrow$)\\
\midrule
Ground-Truth (SLT/American) &76.48{\tiny $\pm$ {3.82} } & 9.56{\tiny $\pm$ {1.32} }\\
MacST (SLT/American) &70.95{\tiny $\pm$ {4.07} } & 10.78{\tiny $\pm$ {1.41} }\\
\addlinespace[0.1em]\hdashline\addlinespace[0.3em]
Ground-Truth (ASI/Hindi) &85.17{\tiny $\pm$ {1.87} } & 67.67{\tiny $\pm$ {2.60} }\\
Ground-Truth (TNI/Hindi) &81.29{\tiny $\pm$ {2.76} } & 70.74{\tiny $\pm$ {2.40} }\\
MacST (SLT/Hindi) &69.51{\tiny $\pm$ {3.99} } & 51.61{\tiny $\pm$ {3.02} }\\
MacST (ASI/Hindi) &82.12{\tiny $\pm$ {2.36} } & 73.61{\tiny $\pm$ {2.51} }\\
MacST (TNI/Hindi) &79.64{\tiny $\pm$ {2.82} } & 77.35{\tiny $\pm$ {2.66} }\\
\midrule
Ground-Truth (SLT/American) &66.84{\tiny $\pm$ {3.45} } & 6.90{\tiny $\pm$ {1.07} }\\
MacST (SLT/American) &70.37{\tiny $\pm$ {3.52} } & 8.56{\tiny $\pm$ {1.40} }\\
\addlinespace[0.1em]\hdashline\addlinespace[0.3em]
Ground-Truth (HKK/Korean) &75.28{\tiny $\pm$ {2.55} } & 39.08{\tiny $\pm$ {2.46} }\\
Ground-Truth (YDCK/Korean) &78.84{\tiny $\pm$ {1.87} } & 32.90{\tiny $\pm$ {2.10} }\\
MacST (SLT/Korean) &58.47{\tiny $\pm$ {4.85} } & 77.63{\tiny $\pm$ {2.33} }\\
MacST (HKK/Korean) &63.22{\tiny $\pm$ {4.06} } & 83.40{\tiny $\pm$ {1.67} }\\
MacST (YDCK/Korean) &63.87{\tiny $\pm$ {4.36} } & 83.44{\tiny $\pm$ {1.67} }\\
\bottomrule
\end{tabular}
}
%\vskip-1.0em
\end{table}

% \subsection{Accent Conversion (American $\rightarrow$ Hindi)}
\subsection{Accent Conversion (American-to-Hindi conversion)}

We developed an accent conversion model to transform American-accented speech into Hindi-accented speech, utilizing an American English speaker's input from CMU-ARCTIC. We trained the models using American-Hindi speech sample pairs. Using MacST, we generated speech samples from 1,132 ARCTIC dataset transcriptions and 4,500 transcriptions from the VCTK dataset. We evaluated two models: the first employed paired data, combining CMU-ARCTIC's ground-truth input with MacST's synthetic Hindi-accented output. The second model added synthetic pairs, including 932 samples (approximately 1 hour) from ARCTIC and 4,500 samples (about 3 hours) from VCTK to test MacST's data augmentation efficacy.

Besides MUSHRA tests, we conducted objective evaluations to assess speech quality, accentedness, and speaker similarity. For speech quality, we measured Word Error Rate (WER). For accentedness, we utilized classification probability and Accent Encoding Cosine Similarity (AECS) difference. For speaker similarity, we used Speaker Encoding Cosine Similarity (SECS).

%Besides MUSHRA tests, we conducted objective evaluations to assess speech quality, accentedness, and speaker similarity.  
Table~\ref{table:result_conversion} presents subjective and objective scores. SECS results confirm the consistent speaker characteristics between the source and converted audio, underscoring our method's effectiveness in maintaining speaker traits.
Other results highlight that accent conversion significantly increased accentedness across all metrics, from ``Ground-Truth (SLT/American)'' to ``AC'' results. Additionally, data augmentation notably enhanced the conversion results in speech quality and accentedness, underscoring the effectiveness of our method in training accent conversion models.
%To validate our findings, we calculated the speaker similarity between SLT and other American speakers using ground-truth audio from CMU-ARCTIC (BDL, RMS, and CLB). SECS values of 0.529, 0.535, and 0.709 for BDL, RMS, and CLB, respectively

%- if speaker similarity result is not good in wavlm and wespeaker, we can mention this.
%    - We had a small analysis of these models' accent dependency and figured out that Resemblyzer is the most accent-independent.

%\subsection{Accent Change}
%\textcolor{blue}{finish here after experiments}
%Here, try to change the non-native accent to another non-native accent.
%another language: Hindi, Korean, or Japanese but Hindi and Korean might be better since these are included in L2-ARCTIC. 

\section{Conclusion}

We introduce MacST, an approach for generating parallel datasets for accented speech via text transliteration. Our method employs Large Language Models (LLMs) to derive transliterated texts, which are then input into multilingual Text-to-Speech (TTS) models to synthesize accented English speech. Dataset analysis confirms MacST's capacity to amplify accents in native and non-native English speakers by highlighting the absence of certain English phonemes in native non-English languages. Both subjective and objective evaluations of the accent conversion validate the efficacy of our method in training accent conversion models.

\newpage
{\footnotesize
\bibliographystyle{IEEEbib}
\bibliography{refs}
}

\end{document}